\documentclass[times, twocolumn]{aastex631}

\begin{document}
\title[Is the Pal 14 GC captured by the MW?]{Has the Palomar 14 globular cluster been captured by the Milky Way? }
\correspondingauthor{Akram Hasani Zonoozi}
\email{a.hasani@iasbs.ac.ir}

\author[0000-0002-0322-9957]{Akram Hasani Zonoozi}
\affiliation{Department of Physics, Institute for Advanced Studies in Basic Sciences (IASBS), PO Box 11365-9161, Zanjan, Iran}
\affiliation{Helmholtz-Institut f\"ur Strahlen-und Kernphysik (HISKP), Universit\"at Bonn, Nussallee 14-16, D-53115 Bonn, Germany}

\author{Maliheh Rabiee}
\affiliation{Department of Physics, Institute for Advanced Studies in Basic Sciences (IASBS), PO Box 11365-9161, Zanjan, Iran}

\author[0000-0002-9058-9677]{Hosein Haghi}
\affiliation{Department of Physics, Institute for Advanced Studies in Basic Sciences (IASBS), PO Box 11365-9161, Zanjan, Iran}

\author[0000-0002-7301-3377]{Pavel Kroupa}
\affiliation{Helmholtz-Institut f\"ur Strahlen-und Kernphysik (HISKP), Universit\"at Bonn, Nussallee 14-16, D-53115 Bonn, Germany}
\affiliation{Faculty of Mathematics and Physics, Astronomical Institute, Charles University in Prague, V Hole\v{s}ovi\v{c}k\'ach 2, CZ-180 00 Praha 8, Czech Republic\\}
\begin{abstract}

We examine a new scenario to model the outer halo globular cluster (GC) Pal 14 over its lifetime by performing a comprehensive set of direct N-body calculations. We assume Pal 14 was born in a now detached/disrupted dwarf galaxy with a strong tidal field. Pal 14 evolved there until the slope of the stellar mass function (MF) became close to the measured value which is observed to be significantly shallower than in most GCs. After about 2-3 Gyr, Pal 14 was then captured by the Milky Way (MW). Although the physical size of such a cluster is indistinguishable from a cluster that has lived its entire life in the MW, other parameters like its mass and the MF-slope, strongly depend on the time the cluster is taken from the dwarf galaxy. After being captured by the MW on a new orbit, the cluster expands and eventually reaches the appropriate mass and size of Pal 14 after 11.5 Gyr while reproducing the observed MF.   These simulations thus suggest that Pal 14 may have formed in a dwarf galaxy with a post-gas-expulsion initial half-mass radius and mass of $r_h=7$ pc and $8<M/10^4<10 $ M$_{\odot}$, respectively,  with a high degree of primordial mass segregation.

\end{abstract}
\keywords{method: numerical - globular clusters: Palomar14 - dwarf galaxy}

\date{Accepted XXX. Received YYY; in original form ZZZ}


\section{Introduction} \label{sec:intro}

Globular clusters (GCs)  represent an interesting family of stellar systems in which some fundamental dynamical processes have taken place on time scales shorter than the age of the universe. GCs are among the most useful objects for studying stellar astrophysics, consisting of many stars of essentially the same age and chemical composition. Moreover, GCs represent the oldest collisional systems, unique laboratories for learning about two-body relaxation, mass segregation from equipartition of energy, stellar collisions, stellar mergers, and core collapse. About 160 GCs \citep{Harris2010}, whose distances from the Galactic center range from 0.5 to 125 kpc, are distributed in and around the Milky Way (MW). More than 50\% of the GCs are known to exist in at least four satellite dwarf galaxies that are orbiting around the MW: the Large Magellanic Cloud (LMC), Small Magellanic Cloud (SMC), Sagittarius, and Fornax \citep{Forbes2010}. If a dwarf galaxy interacts with the MW its GCs may be incorporated into the MW’s GC population. The Sagittarius dwarf galaxy is one of the benchmark examples interacting with our Galaxy \citep{Ibata1994, Law2010}. It is also thought that a fraction of the MW GCs are accreted systems, while the remaining ones formed in the early phase of Galaxy formation \citep{Forbes2010}. Comparison of the MW GCs with those in satellite galaxies shows that all 30 of the young halo clusters as well as $10-12$ old halo clusters may have originated in dwarf galaxies leading to the estimate that the MW has encountered approximately 7 dwarf spheroidal type galaxies \citep{Mackey2004}.  Therefore, Galactic GCs might have formed in situ in the Galaxy or in a dwarf galaxy to be captured by the MW \citep{Rostami2023}. The young halo GCs are associated with the vast polar structure \citep{Pawlowski2014} and are therefore related to the satellite galaxies populating this structure.

The outer-halo globular clusters such as Pal 4 and 14 have large half-light radii compared to the clusters in the inner part of the MW \citep{Harris2010}.  Three main mechanisms have been suggested to explain the large radii of extended star clusters: i) these clusters might be born extended \citep{Elmegreen2008}; ii) they might be formed during the merging of two or more clusters \citep{Fellhauer2002, Assmann2011, Bruns2011}; iii) they were born compact and later expanded due to a top-heavy initial mass function (IMF), due to the environmental evolution such as strong tidal shocks or dynamical heating through stellar-mass black holes (BHs) \citep{Mackey2004, Miholics2014, Miholics2016}. Additionally, interactions between stars and BHs can lead to the expulsion of stars from a cluster. This can have an impact on the size and structure of the cluster. In some cases, the retention of BHs may result in the formation of an extended cluster, such as Pal 5, as demonstrated by \citet{Gieles2021}.

During the gravitational interaction between two galaxies, the cluster undergoes a change in potential which could cause a change in the cluster’s characteristic parameters. For example, a large half-mass radius of a GC is sometimes assumed to be evidence that a cluster may have been captured from a dwarf galaxy, as was suggested for Pal 14 \citep{Frank2014}. \cite{Miholics2014} performed N-body simulations of a cluster that forms in a dwarf galaxy and is then acquired by the MW to investigate how a cluster’s structure is affected by interacting galaxies. They found that the cluster’s half-mass radius will respond quickly to this change in potential. When the cluster is placed on an orbit in the MW with a stronger tidal field the cluster experiences a sharp decrease in size in response to increased tidal forces. However, later \cite{Miholics2016} found out that ultimately such clusters quickly become the same size as a cluster born in the MW on the same orbit.

Using N-body simulations, \cite{Bianchini2015} showed that if a cluster forms in the centre of a dwarf galaxy, it undergoes a sizable expansion, during the drastic variation of the external tidal field due to the capture process. However, the expansion is not enough to explain the observed extended structure. They conclude that an acquired origin of extended GCs is unlikely to explain their large spatial extent. In this regard, \cite{Webb2017} found that if the clusters experience the strong tides only during the early evolutionary stages, right after the transition to weak tides, they can be more extended and less massive than clusters that experienced only a weak tidal field. Moreover,  \cite{Bianchini2017} analyzed the effects imprinted on the internal kinematics of a  captured GC and showed that clusters will lose any kinematical signature of their original environment in a few median two-body relaxation times. 

Another initial condition that is effective in the final size of the cluster is primordial mass segregation (PMS) which refers to the phenomenon where more massive stars are preferentially located in the central regions of a star cluster, while lower-mass stars are more widely distributed throughout the cluster. Star clusters may form mass segregated at birth \citep{Pavlik2019}. Many theoretical models of star cluster formation predict the presence of PMS. These models typically involve the collapse of a dense cloud of gas and dust, which eventually forms a star cluster.

Overall, PMS is an important factor to consider when studying the evolution of star clusters, as it can have significant effects on a variety of physical processes within the cluster \citep{Heggie2003}. In particular, it can impact the rate of stellar interactions and the overall dynamical evolution of the cluster \citep{Vesperini2009, Baumgardt2008}. Mass segregation can also affect the rate of stellar collisions and mergers in a star cluster. In a mass-segregated cluster, there are more opportunities for high-mass stars to collide with each other, leading to the formation of even more massive stars. This can potentially result in the formation of intermediate-mass black holes, which could have important implications for the overall evolution of the cluster \citep{Rose2022}.

Regardless of the mechanism producing mass segregation, the presence of primordial (or early) mass segregation significantly affects the global dynamical evolution of star clusters. In tidally limited clusters, PMS leads to a stronger expansion, and hence a larger flow of mass over the tidal boundary. PMS may therefore help to dissolve clusters more rapidly \citep{Haghi2014, Haghi2015}. Tidally underfilling clusters, however, can survive this early expansion and have a lifetime similar to that of unsegregated clusters, as demonstrated by \citet{Vesperini2009}. The authors furthermore showed that, as the degree of initial mass segregation increases, so does the strength of the initial cluster expansion. Similarly, \citet{Mackey2007, Mackey2008} demonstrated that the stronger early expansion of mass-segregated clusters, along with the subsequent dynamical heating from a population of stellar mass black holes, can explain the radius–age trend observed for massive clusters in the Magellanic Clouds \citep{Gouliermis2004}. The degree of primordial, or early, mass segregation is, therefore, a crucial parameter in the modeling of GCs.

The photometric and spectroscopic investigations of outer-halo GCs, such as Pal 4 and Pal 14, show large half-light radii of  $20-30$ pc, flat stellar mass functions depleted in low-mass stars, and clear signatures of mass segregation within their centers \citep{Jordi2009, Frank2012, Frank2014} which is dynamically unexpected since their present-day two-body relaxation times are larger than a Hubble time. One way of solving this apparent discrepancy is if these GCs were born compact, and the observed mass segregation happened in the early evolution of the clusters \citep{Zonoozi2011, Zonoozi2014}. Moreover, Pal 4 is most likely orbiting on an eccentric orbit with an eccentricity of $ e\approx0.9 $ and pericentric distance of $ R_p\approx5 $ pc \citep{Zonoozi2017}. In all of these works, it is assumed that the cluster was born in the MW, but these GCs might be formed as star clusters in dwarf galaxy satellites. However, no study has been done to test this scenario.

In this paper, we examine how the stellar MF slope of a GC would change if it is captured by the MW. We perform N-body simulations of a cluster that starts in a dwarf galaxy potential and is then instantaneously switched to MW potential to emulate a capture process.  The important question we aim to answer is whether we still need PMS to explain the flattening of the MF in Pal 14 in the context of acquired GCs. The paper is organized as follows: We provide a brief overview of the theoretical and observational evidence for PMS in Sec. 2. Section 3 presents the observational data for Pal 14. In Section 4, we describe the setup of the N-body models. This is followed by a presentation of the results of the simulations and comparison with observed parameters in Section 5 and we finally summarize our work in Section 6.

\section{Theoretical and Observational Evidence for PMS}

PMS is possible as a result of the star formation process \citep{Zinnecker1982, Murray1996, Klessen2001, Bonnell2001, Bonnell2006} where the massive stars tend to form in the centers of the star-forming regions because the higher accretion rates in the central parts or as a result of competitive accretion \citep{Larson1982, Murray1996, Bonnell1997}. Such a primordial state has been used to explain very high levels of mass segregation (MS) in young systems when solely dynamical processes are not fast enough to explain the observed MS \citep{Bonnell1998, Raboud1998}. Moreover, a large number of young clusters with ages significantly smaller than the time needed to produce the observed mass segregation by two-body relaxation alone show a significant degree of mass segregation, which would probably be primordial and imprinted in a cluster by the star formation processes \citep{Pavlik2019}. For example, studies of the Orion Nebula Cluster (ONC), a relatively young cluster with an estimated age of 1-2 million years, have found that the most massive stars are preferentially located in the central region of the cluster. This suggests that some degree of mass segregation was present from the very beginning of the cluster's formation. In addition, observations of other young clusters, such as Pleiades, NGC 3603,  Westerlund 1, and Westerlund 2  have also found evidence for MS \citep{Converse2010, Zeidler2017, Lim2013, Baumgardt2022}.

MS has been detected even in embedded star clusters \citep{Lada1996, Hillenbrand1997, Hillenbrand1998, Bonatto2006,  Er2013, Plunkett2018}, this means that the most massive stars are preferentially concentrated (not necessarily in the center). There is also observational evidence of PMS in several young Galactic and Magellanic Cloud star clusters with ages shorter than the time needed to produce the observed segregation via two-body relaxation \citep{Hillenbrand1997, Bonnell1998, deGrijs2002, Sirianni2002, Stolte2006, Sabbi2008, Allison2009}. These star clusters may have formed out of many smaller star-forming clumps. In each clump, rapid mass segregation may have occurred, sending the most massive stars to the core of each clump. \citet{McMillan2007} showed that when such clumps merge, they will quickly form a virialized cluster, but the mass segregation of the clumps is largely preserved. Under MW molecular cloud conditions, young clusters are unlikely to merge through \citep{Mahani2021}. Alternatively, the PMS in young star clusters could result from star formation feedback in dense gas clouds \citep{Murray1996}, or due to competitive gas accretion and mutual mergers between protostars \citep{Bonnell2001}. Another evidence for PMS was proposed by \citet{Marks2008} to explain the flattening of the stellar mass functions seen in some low-concentration GCs, together with the correlation between the slope of the stellar mass function and the cluster concentration that had been discovered by \citet{DeMarchi2007}. \citet{Baumgardt2008} found that clusters with PMS lose their low-mass stars at a higher rate than non-segregated ones if evolving in a strong external tidal field because low-mass stars move in the outer parts of the cluster, where the tidal field easily removes them. This effect is enhanced if residual gas removal is taken into account because the sudden drop in the cluster potential as a result of gas expulsion leads to preferential loss of low-mass stars moving at large radii \citep{Marks2008}. Given the observational evidence for PMS in young star clusters as well as old diffuse GCs, we conclude that at least some, but possibly all, GCs must have started with PMS.

Models of star cluster formation typically involve the collapse of a dense cloud of gas and dust, eventually forming a star cluster. During the early stages of this process, the more massive stars are expected to form where there are higher accretion rates.  As a result, these more massive stars tend to be located in the central regions of the cluster, while lower-mass stars are distributed more widely throughout the cluster. Overall, the degree of PMS predicted by these models depends on a variety of factors, including the initial conditions of the cloud, the efficiency of star formation, and the strength of any feedback processes that may be present. However, in general, many models predict at least some degree of mass segregation in the early stages of star cluster formation.  Current existing data such as ALMA observations of the Serpens South star-forming region appear to suggest that stellar protoclusters may be completely mass-segregated at birth \citep{Plunkett2018}.  In the case of the ONC, using numerical N-body models of star clusters, \citet{Pavlik2019} showed that initially, perfectly mass-segregated models seem to be more consistent with the observed cluster.



%
%

\section{OBSERVATIONAL DATA}


To compare the results of our modeled GCs that form in a dwarf satellite galaxy, with the present-day properties of Pal 14, we used observational data by \cite{Hilker2006} and \cite{Jordi2009}. They have obtained a deep color-magnitude diagram (CMD) of Pal 14, using  Wide Field Planetary Camera 2 (HST/WFPC2) data from the Hubble Space Telescope archive. Analyzing  Pal 14's CMD to about 4 mags below the main sequence turnoff point, they have determined the cluster's stellar mass function down to about $0.49 \mathrm M_{\odot}$. 

Using Dartmouth isochrones \citep{Dotter2007} and adopting a metallicity of $[\mathrm {Fe}/\mathrm H]=-$1.50 \citep{Harris1996, Harris2010}, the age of Pal 14 is estimated to be $(11.5 \pm 0.5) $ Gyr by \citet{Jordi2009}. However previous works have derived a younger age of $\approx 10 $ Gyr \citep{Sarajedini1997, Hilker2006, Dotter2007} with a larger offset to the ages of typical halo GCs.  Pal 14's distance from the Sun is derived to be about $71\pm 1.3 $ kpc from a CMD and the isochrone fit \citep{Jordi2009}. This is slightly closer to the Sun than the previous estimation of 74.7 kpc given by  \cite{Hilker2006} and 79 kpc by \cite{Dotter2007}. Pal 14 thus counts to the population of young halo GCs and is part of the Vast Polar Structure, a polar plane in which most of the satellite galaxies orbit \citep{Pawlowski2014}. 
The projected half-light radius of Pal 14 is obtained to be 1.28 arcmin \citep{Hilker2006}. Assuming that Pal 14 is at a distance of about 71 kpc, this value corresponds to a projected half-light radius of $\mathrm R_{\mathrm {phl}} = 26.4 \pm 0.5 $ pc and a 3D half-mass radius of about $\mathrm r_{\mathrm h}=35.6 \pm 0.6 $ pc \citep{Zonoozi2011}. 

The mass function of Pal 14 is determined to have a power-law index, or slope of $1.27 \pm 0.44$ for main sequence stars in the mass range of 0.525 M$_{\odot}$ and 0.795 M$_{\odot}$ \citep{Jordi2009}, where a slope of $2.35$ is the Salpeter index. The data of this mass range which is considered to estimate the MF-slope has a completeness factor greater than 0.5.  The total mass of stars within the half-light radius has been estimated at M = $ 2200\pm90$ M$_\odot $, within the mass range 0.525 - 0.795M$_ \odot $. With the measured slope for stellar masses between 0.1M$ _\odot $ and 0.5M$ _\odot $ using a Kroupa-like IMF, \cite{Jordi2009} estimated the total mass of Pal 14 of M$_{\mathrm tot}\approx12000$ M$_\odot $, without taking into account stellar remnants.




\section{N-BODY MODELS} \label{sec:model}

This section discusses our choices for initial mass, half-mass radius, PMS,  and orbital parameters of the modeled star clusters. We used Sverre Aarseth’s direct N-body code NBODY6 \citep{Aarseth2003, Nitadori2012} on a desktop workstation with Nvidia 690 Graphics Processing Units at the Institute for Advanced Studies in Basic Sciences(IASBS) to model the dynamical evolution of Pal 14 over its entire lifetime.

The initial configuration for all model clusters was set up using the publicly available code \textsc{McLuster}\footnote{https://github.com/ahwkuepper/mcluster} \citep{Kuepper2011}. 
Initially, the distribution of stellar positions and velocities follow that of a Plummer density profile \citep{Plummer1911, Aarseth1974} out to a cut-off radius of two times the cluster’s tidal radius, $r_t$, and star velocities are set so that the cluster is in virial equilibrium with the surrounding tidal field. 

For each model, we adopt the canonical initial mass function (IMF) to randomly generate the initial distribution of stellar masses between $ m_{min}$ = 0.08 $ M_\odot $ and $ m_{max} $ = 100 $ M_{\odot} $ in the form of \citep{Kroupa2001}

\begin{eqnarray}
\,\,\,\,\frac{dN}{dm}\propto m^{-\alpha},\,\,\,\, 
\,\left\{
      \begin{array}{ll}
\, \alpha_1= 1.3  &\,\,\, 0.08M_\odot\leq m < 0.5M_\odot,  \\
\, \alpha_2= 2.3   &\,\,\, 0.5M_\odot\leq m < 100M_\odot
       \end{array}
        \right. \label{MF}
\end{eqnarray}

It should be noted that the sampling method of stars is important for all star clusters, but it is particularly crucial for low-mass clusters where individual stars play a more significant role in shaping the cluster's evolution. Low-mass star clusters have fewer stars compared to high-mass clusters, which means that each star has a greater impact on the overall characteristics and evolution of the cluster.  In contrast, high-mass star clusters have a larger number of stars, which can provide a more robust statistical sample.  In cases where we only change the cluster's orbital parameters, we use the same generated cluster. 

The evolution of individual stars, all assumed to have a metallicity of $0.001$, follows the single and binary stellar evolution algorithm from the ZAMS through remnant phases by using the SSE/BSE routines and analytical fitting functions developed by \cite{Hurley2000} and \cite{Hurley2002}. The BHs and NSs receive a kick velocity drawn from a Maxwellian distribution with a one-dimensional dispersion of $\sigma_{\text{kick}} = 190 $km/s \citep{Hansen1997} and almost all of them are kicked out of the cluster immediately after formation. It should be noted that SSE/BSE routines, implemented in NBODY6, underestimate the remnant's masses, particularly at intermediate and low metallicities \citep{Belczynski2010}. Moreover, different implementations of the envelope mass fraction that falls back onto the remnant and pair-instability supernova treatment can affect the final kick velocity \citep{Belczynski2008, Belczynski2016, Banerjee2020}  and possibly lead to the reduction of the BH natal kick velocities \citep{Pavlik2018}. This, however, does not have any significant effect on our present calculations as we have assumed the retention fraction to be zero for the BHs and NSs. However, in the future, we will focus on examining the impact of different retention fractions. Having a compact population of massive BHs in the cluster core will likely influence the early dynamics of the system, potentially affecting the final half-mass radius and stellar mass function slope (Zonoozi et al. in prep.) Moreover, we assume that any residual gas from the star formation process has been removed. Our model is thus a post-gas-expulsion cluster probably about four times larger than when it was in its embedded phase \citep{Brinkmann2017}.  

The primordial binary fraction plays a crucial role in shaping the dynamical and evolutionary properties of star clusters, impacting their overall structure, dynamics, and mass function \citep{Pavlik2020, Motherway2024}. Binaries tend to sink towards the center of the cluster due to dynamical friction, leading to mass segregation. This can affect the overall mass function of the cluster by concentrating more massive stars in the core. Moreover, binary stars can undergo interactions that can release energy into the cluster. This energy can lead to the evaporation of low-mass stars, effectively altering the mass function of the cluster.  Since the dynamical effects from primordial binaries are not expected to be significant for such an extended star cluster as Pal 14 \citep{Kroupa1995a, Kroupa1995b}, we do not consider primordial binary stars for the sake of computability.  It has already been shown by \citet{Zonoozi2011}  that adding a 5\% primordial binary fraction has no significant effect on the evolution of an extended GC such as Pal 14. They found that the binary fraction of Pal 14 stays almost the same and gives the final fraction over its entire lifetime due to the cluster’s extremely low density. A small fraction of binaries can form via three-body interactions throughout the simulations \citep{Kroupa1995a, Kroupa1995b}.

The code \textsc{McLuster} allows initializing any degree of PMS to all available density profiles using the routine described in \cite{Baumgardt2008}. This routine ensures maintenance of the desired mass density profile and virial equilibrium when increasing the degree of mass segregation. For a fully segregated cluster ($S = 1$), the most massive star occupies the orbit with the lowest energy, the second most massive star occupies the orbit with the second lowest energy, and so forth.  In this paper, some modeled clusters are assumed to be not initially mass segregated ($S = 0$), but to explain the observational values, we compute several clusters with PMS.

To study how clusters with different initial masses and sizes evolve in the tidal fields, we consider different models with a total initial stellar mass in the range $ 2.8\times10^4<M/M_\odot\leq 1.0\times10^5 $ and initial 3D half-mass radii are in the range $5-8$ pc.  We assume the cluster is born with the birth radius \citep{Marks2012} and gas expulsion expands to the initial radius used here.  We compute a set of 32 N-body models to find the initial conditions that best reproduce the observations of Pal 14.

To set up a strong external compressive tidal field, we place our model clusters at different distances from the center of a dwarf galaxy (in the range of $R_{\{g,dw\}}=2-4$ kpc) with a mass of $10^9 M_\odot$.  Clusters spend a fraction of their lifetime in a compressive tidal field.  More specifically, we allow different models to evolve in a compressive tidal field of the dwarf galaxy over 1.5 to 10 Gyr.  We then move each model cluster to a range of extensive tidal fields for the rest of their lifetime. Extensive tidal fields are set up by placing the cluster in a Milky Way-like tidal field at Galactocentric distances of 71 kpc on circular orbits. In six models, clusters evolve on eccentric orbits. We followed the evolution of the clusters for 11.5 Gyr, during which they typically lose about 70–85 percent of their mass into the Galactic halo.   Compared to the models evolving on a circular orbit, we chose a smaller initial radius and a larger initial mass for models on eccentric orbits. This is because of the enhanced mass-loss rate driven by tidal shocks as the cluster passes through perigalaction (in addition to the mass-loss by stellar evolution and from dynamical evolution).

We used a three-component Milky Way-like tidal field which is made up of a bulge, a disc, and a phantom dark matter halo potential \citep{Lughausen2015} that is scaled so the circular velocity at 8.5 kpc is 220 km/s. The bulge is modeled as a central point mass:
\begin{equation}
    \phi_b(r)=-\frac{GM_b}{r},
	\label{eq:3-1}
\end{equation}
where $ M_b$ is the mass of the bulge component. The potential of  the disc is approximated by the  Miyamoto \& Nagai potential \citep{Miyamoto1975};
\begin{equation}
    \phi_d(X,Y,Z)=-\frac{GM_d}{\sqrt{X^2+Y^2+(a+\sqrt{Z^2+b^2})^2}},
	\label{eq:3-2}
\end{equation}
where $a$ is the disc scale length, $b$ is the disc scale height, and $ M_b$ is the total mass of the disc component. We used values of $ a =4$kpc and $ b=0.5 $kpc \citep{Read2006}, while for the disc and bulge masses, we adopted $ M_d = 5\times10^{10}M_\odot$ and $ M_b=1.5\times10^{10}M_\odot $ respectively, as suggested by \cite{Xue2008}. The phantom halo is represented by a logarithmic potential given by:
\begin{equation}
        \phi_H= \frac{1}{2} V_{\rm circ}^2 \log_{10} (R_{gal}^2+R_{\rm circ}^2).
	\label{eq:3-3}
\end{equation}
where $R_{\rm gal}$ is the Galactocentric radius. $V_{\rm circ}=220~\rm kms^{-1}$ is the circular velocity of the body at radius $ R_{\rm circ}=8.5~\rm kpc $ in the disk plane in the combined gravitational potential provided by the bulge, disc, and halo \citep{Xue2008}. 

In this work, all model clusters except two are initially assumed to orbit in a circular orbit of various radii around a point mass dwarf galaxy with $M_{dw}$ = $10^9M_\odot $. When the IMF $ \alpha_2 $ parameter reaches $ \alpha_2\approx1.4 $, which we call the flattening time ($ t_\alpha $), we take the cluster out of the dwarf galaxy and place it at a distance of $ 71 \rm kpc $ of the MW galaxy. The choices for the initial conditions of all models and the key results of the simulated clusters are summarized in Table  \ref{table_1}.



\section{Results}
\label{sec:Res}

In this section, we present the results of our numerical simulations that start from a set of initial conditions with several parameters that can be varied to find the best-fitting model. We first generally examine how the MF slope of a GC would change if it starts in a dwarf galaxy potential and is then instantaneously switched to the MW potential on circular and eccentric orbits to emulate a capture process. Hereby we choose real orbital data of Pal 14 using GAIA DR3 to check this scenario by finding the best-fit simulated model.  The modeled clusters are different in initial mass,  half-mass radius, orbital distance from the center of the dwarf galaxy, and degree of PMS. We keep only bound stars and remove stars with $ r>2r_t$ from the computations, where $r_t$ is the instantaneous tidal radius.  To find the best-fitting model for Pal 14, we compare the results of N-body simulations with observational data described in Section 2 and listed in the last row of Table~\ref{table_1} in terms of the present-day total mass $(M_{\rm tot}=12000M_\odot)$,  half-mass radius $(r_{\rm h,f}=35.6\pm0.6 \rm pc)$, and the global MF-slope over the stellar mass range $0.5-0.8 M_\odot $ within the half-mass radius $ (\alpha_2=1.27 \pm 0.44) $.  The main initial properties of our models and the key results of the simulated clusters, that is, final radius, total mass, and global MF-slope are listed in Table~\ref{table_1}.  To study the influence of PMS on the present-day properties of star clusters in detail, we performed two sets of models, with and without PMS. Both will be discussed in the following.

 \begin{figure}
   
    \centering
   \includegraphics[width=8CM]{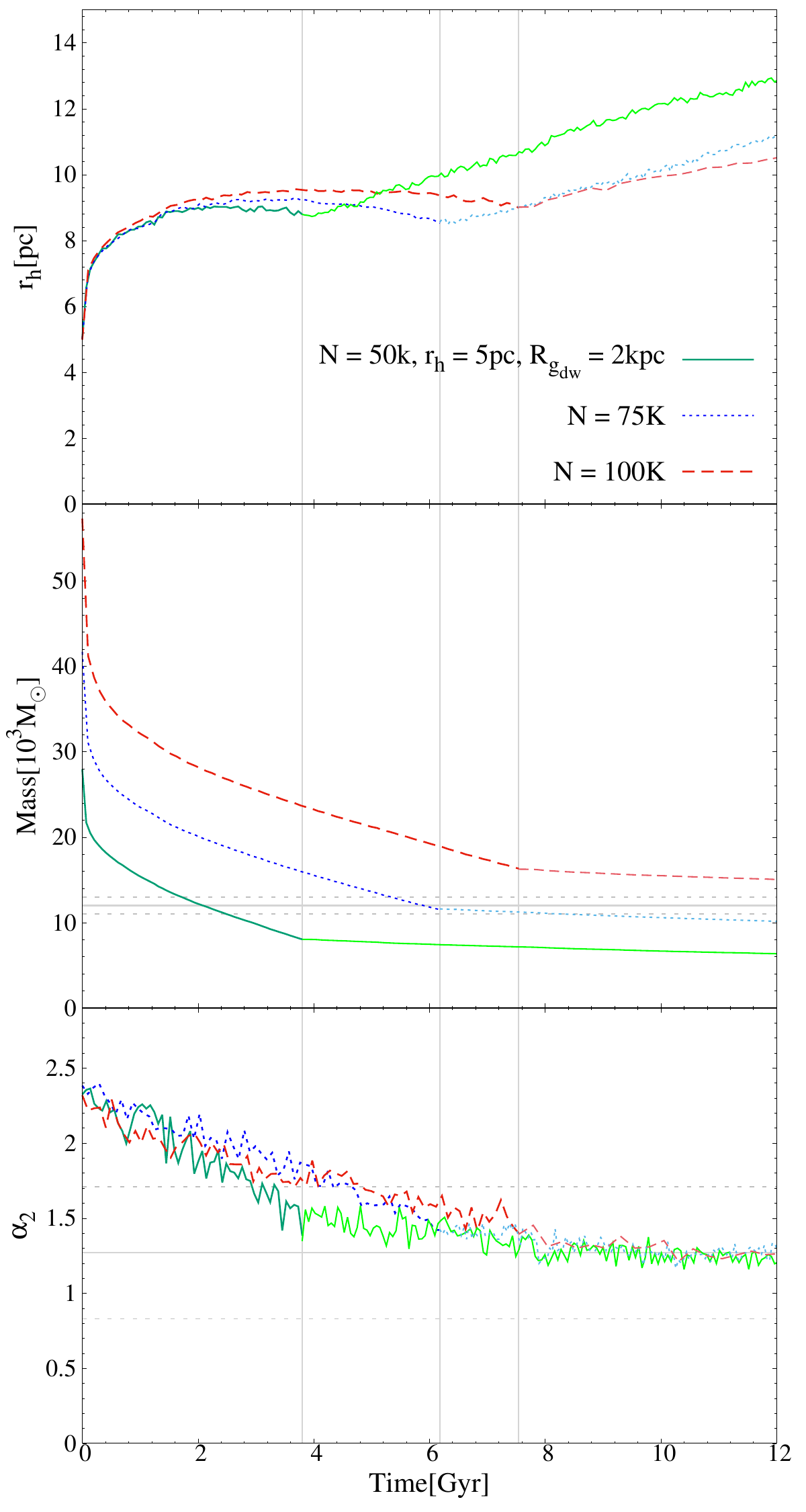}
\caption{Evolution of the half-mass radius (top), the total mass (middle), and $\alpha_2$ (bottom) of different model clusters without PMS ($S=0$). All models start with the same 3D half-mass radius of 5 pc but different numbers of particles: $N = 5\times10^4$ (green solid lines, Model 2), $N = 7.5\times10^4$ (blue dotted lines, Model 7), and $N = 10^5$ (red dash lines, Model 9). In all models, firstly the clusters evolve around a point mass dwarf galaxy with $ M_{dw}=10^{9}M_\odot $ at an orbital distance of 2 kpc, and at the switching time, they are placed on a circular orbit in the MW at the Galactocentric distance of 71 kpc. The $ \alpha_2 $ parameter is calculated for stars in the mass range $0.5< M/M_\odot < 0.8$ and inside the half-mass radius. Vertical lines show the corresponding $t_{sw}$ to each model (Table 1). The solid horizontal lines represent the observed values, and the dashed horizontal lines indicate the 1$-\sigma$ errors.  The observed value of $r_h=35.6\pm0.6$ pc is not shown in the top panel due to its large value.} The very early drop of the mass and increase of the half-mass radius is due to the stellar evolution. 
\label{fig:1}
\end{figure}

\begin{figure}

    \centering
   \includegraphics[width=8CM]{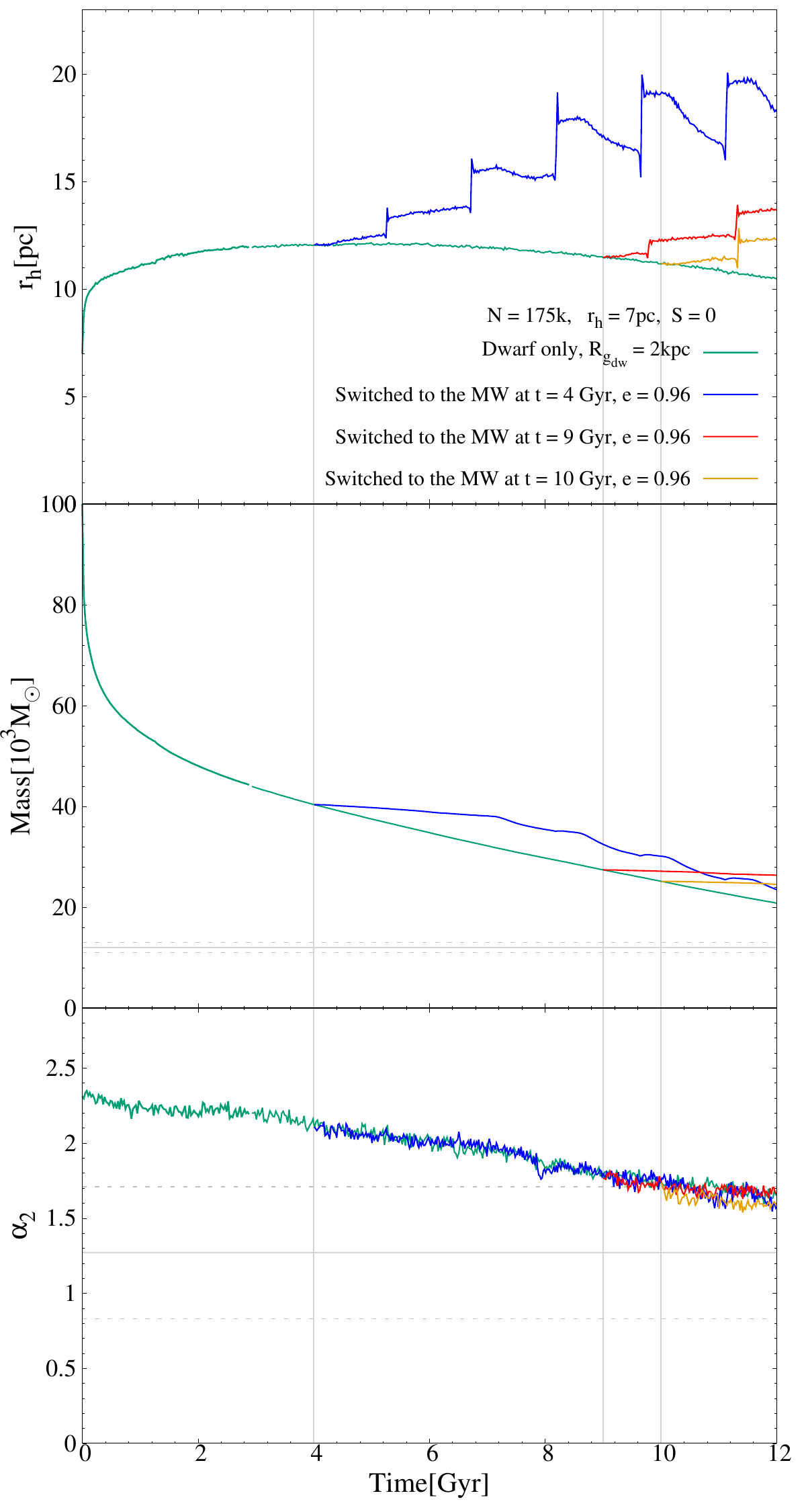}

\caption{Evolution of the half-mass radius (top), the total mass (middle) and $\alpha_2$ (bottom) of the clusters without PMS and with $N=1.75\times10^5$ particles and an initial 3D post-gas expulsion half-mass radius of 7 pc.  The cluster evolves around a point mass dwarf galaxy, with $ M=10^9M_\odot$ in an orbital distance $R_{\{g,dw\}}=2 \rm kpc$ (green solid lines), and has been switched from dwarf galaxy potential to the Milky Way potential in an eccentric orbit with $ e\approx0.95 $  at three different times of $t_{sw}= 4, 9, 10 \rm Gyr$. The horizontal lines show the observed values. Vertical lines show the corresponding $t_{sw}$ to each model (Table 1).}

\label{realorbit}
\end{figure}

\subsection{Models without PMS} \label{sec:withoutPMS} 


All clusters in this section (i.e., models 1 to 11 in Table~\ref{table_1}) are initially non-segregated, starting with a canonical IMF \citep{Kroupa2001}. Fig.~\ref{fig:1} depicts the evolution of the total bound mass, 3D half-mass radius, and MF-slope as a function of time for some selected models with different initial numbers of stars but the same initial half-mass radius orbiting at the same orbital distance from the center of the dwarf galaxy ($ R_{\{g,dw\}}=2 $ kpc, Models 2, 7 and 9 of Table~\ref{table_1}). The horizontal lines show the present-day total bound mass,  3D half-mass radius, and MF-slope of Pal 14. The sharp change indicates the moment the cluster moves from the dwarf to the MW potential. During the evolution in a dwarf galaxy, most of the stars beyond the perigalactic tidal radius are stripped away from the cluster as a result of a strong tidal field, leading to a quick decrease in bound cluster mass. As the cluster moves to the MW potential at a Galactocentric distance of 71 kpc, the number of cluster stars changes fairly smoothly.

Moreover, Fig.~\ref{fig:1} shows the mass function of more massive clusters needs more time to be flattened, and after $t_{sw}$, when the clusters switched to the MW, the $ \alpha_2 $ parameter changes slowly. The half-mass radius of clusters with different initial masses is almost the same at their switching time. After the switching time ($t_{sw}$), model 2 (green solid line), which is transferred to the MW sooner and has more time to evolve in the MW, shows a larger final half-mass radius, but it is still below the observed value of $r_{\rm h,f}=35.6\pm0.6 \rm pc$.

Increasing $M_0$ or/and $R_{\{g,dw\}}$ leads to a later flattening of the mass function slope. However, the increased mass loss due to a larger initial half-mass radius of the cluster leads to severe depletion of low-mass stars and hence a quicker change of the mass function. Models  10 and 11 appear to have not reached $\alpha=1.4$ within the dwarf galaxy till the end of the simulations ($> 14 \rm Gyr$). As can be seen from the upper section of Table 1, the final half-mass radius in all our computed models is less than 19 pc, and therefore lower than the observational value. However,  many models can reproduce Pal 14’s total mass and MF-slope, but none of them can reproduce the observed value of the half-mass radius of Pal 14 which has an unusually large size. As shown in Table~\ref{table_1}, even if we start the simulation with $ r_h$= 8 pc, the half-mass radius of the cluster at $ t=11.5 \rm Gyr$ is far from the observed value. Hence, we conclude that the models starting without PMS and placed on a circular orbit in the MW  are unable to reproduce the observed 3D half-mass radius of Pal 14 even when accounting for statistical errors. 

\subsubsection{Evolution on realistic orbit}

The cluster orbits in the dwarf galaxy and is then moved to a MW-like potential once its mass function has evolved to a specific value. The cluster's orbit in the Milky Way is chosen based on orbital parameters measured by \citet{Vasiliev2021} and \citet{Bajkova2022} suggesting its orbit is highly eccentric with $e>0.95$. Pal 14's proper motion, derived from Gaia EDR3 astrometry and taking into account spatially correlated systematic errors is $\mu_{\alpha}\cos{\delta}=-4.63\pm0.038$ and $\mu_{\delta}=-4.13\pm0.038$. Given the large eccentricity and small perigalactic distance, the tidal effect in terms of shocks is expected to play an important role in the evolution of the star cluster at least in the outer parts. We calculated 4 models without PMS and with the same initial half-mass radius $(r_h=7 \rm pc)$ consisting of $N=1.75 \times10^5$ stars.  All clusters evolve at the same orbital distance around the dwarf galaxy ($R_{\{g,dw\}}=2$ kpc), but we take the cluster out of the dwarf galaxy and place it onto its realistic orbit around the MW galaxy at different times of 2, 4, 9, and 10 Gyr. 

Fig.~\ref{realorbit}  compares the evolution of the 3D half-mass radius, total mass, and MF-slope for these four models (Models 12-15). According to the overall results summarized in Table 1, it is evident that while the degree of flattening of the mass function slope rises with a longer duration of cluster evolution in the dwarf galaxy, the final radius simultaneously decreases. As can be seen, there is not much difference in the results and it seems that the models without PMS are neither able to accurately match the observed data nor able to account for the observed half-mass radius of Pal 14. Additionally, even with the amount of mass loss is taken into account, the observed mass still doesn't match up with the predictions.

%

%
%
\begin{figure*}
\centering
    \includegraphics[width=18CM]{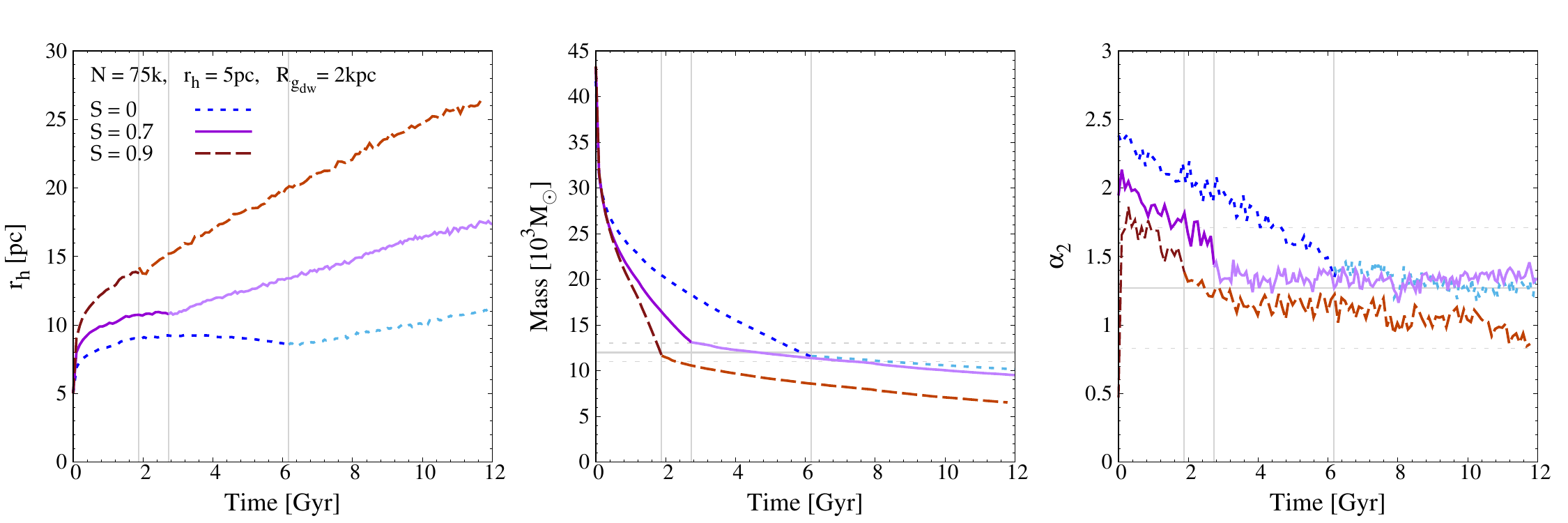}
    \caption{The same as Figure \ref{fig:1}, but for clusters with $N=7.5\times10^4$ particles and different degree of PMS: $S=0$ (dark and light blue dashed lines); $S=0.7$ (purple solid lines) and $S=0.9$ (brown dashed lines). On each curve the dark-colored part corresponds to the evolution period in the dwarf galaxy and the lighter part corresponds to the cluster evolution in the MW. }  
    \label{fig:2}
\end{figure*}
%

\begin{table*}
 \centering
 \caption{Details of the N-body models with and without PMS. Here $S$ is the degree of mass segregation, $N$ is the number of stars; $M_0$ is the total mass of the cluster; $R_{\{g,dw\}}$ is the Galactocentric distance from the dwarf galaxy; $ r_h $ is the initial post-gas-expulsion half-mass radius of the cluster; $ t_{sw} $ is the time at which the cluster switches from the dwarf galaxy potential to the Milky Way potential; $ M_{t_{sw}}$ is the total mass of the cluster at the switching time; $ r_{h,t_{sw}} $ is the cluster's 3D half-mass radius at $ t_{sw} $; $ M_{f} $ is the total mass of the cluster at $ t=11.5 \rm Gyr $; $ r_{h,f} $ is the cluster's 3D half-mass radius at $ t=11.5 \rm Gyr $; $\alpha_{h,f}$ is the value of $ \alpha_2$  at $t=11.5 \rm Gyr$. The best-fitting models that agree within the uncertainties with all observational parameters are highlighted with boldface.  Model 32 is also marginally in agreement with the observation.
 }
 
 \label{table_1}
  \begin{tabular}{ l c c c c c c c c c c c r}
  \hline
  Model &S& N & $M_0$ & $R_{\{g,dw\}}$ & $r_h$ & $t_{sw}$&$\alpha_{t_{sw}}$ & $ M_{t_{sw}}$& $ r_{h,t_{sw}} $ &  $ M_{f}(t=11.5Gyr)$& $ r_{h,f}(pc) $ & $\alpha_{2h,f}$\\ 
 & & $\times10^3$ & $\times10^3 (M_\odot) $&(kpc) & (pc) & (Gyr)& & $\times10^3 (M_\odot) $ & (pc)& $\times(10^3)\pm1100 M_\odot$ & $(\pm2.8)$ &$(\pm 0.13) $\\
\hline
\multicolumn{11}{l}{\textit{ without PMS on circular orbit}} \\
1 &0& 50 & 29.3 &2 & 8 & 1.9 &1.4& 5.5 & 12.7 &3.4 &18.7 & 1.02 \\
2 &0& 50 & 28.2 &  2 & 5 & 3.8 &1.4& 8.0 & 8.8 & 6.4 & 12.6 & 1.17 \\
3 &0& 50 & 29.3  & 3 & 8 & 6.5 &1.4& 5.5 & 14.2 &4.7 &16.4 & 1.46 \\
4 &0& 50 & 28.2 &3 & 5  & 4.6 &1.4& 11.7 & 10.5 & 10.0&13.8 & 1.27 \\
5 &0&50 & 29.3 & 4 & 8  & 8.8 &1.4& 8.1& 16.6 & 7.7& 17.5 & 1.30\\
6 &0& 50 & 28.2 & 4 & 5 & 7.8 &1.4& 11.5 & 11.9 & 10.7 &13.6 & 1.26\\
7 &0& 75 & 41.7 &2 & 5 & 6.2 &1.4& 11.6 & 8.5 & 10.2&10.9 & 1.26\\
8 &0&100 & 57.7 & 2 & 8 & 4.6 &1.4& 9.1 & 13.0 & 7.2&16.6 & 0.95\\
9 &0&100 & 57.5 & 2 & 5 & 7.5 &1.4& 16.3 & 9.0 & 15.1 & 10.3 & 1.28\\
10 &0&100 & 57.5 & 4 & 8 & $>14$ &1.4& - & - & -&-&-\\
11 &0&100 & 57.5 & 4 & 5 & $>14$ &1.4& - & - & -&-&-\\

\hline
\multicolumn{10}{l}{\textit{ without PMS and realistic orbit in the MW}} \\
12 &0& 175 & 100.0 & 2 & 7 & 2 &2.3& 48.0 & 11.7 & 25.0 &  20.8  & 2.13\\
13 &0& 175 & 100.0 & 2 & 7 & 4 &2.1& 40.3 & 12.0 & 25.5 & 19.7 & 1.73\\
14 &0& 175 & 100.0 & 2 & 7 & 9 &1.8& 27.4 & 11.5 & 26.4 &  13.6  & 1.70\\
15 &0& 175 & 100.0 & 2 & 7 & 10 &1.7& 25.1 & 11.2 & 24.6 & 12.2 & 1.60\\ 

\hline
\multicolumn{11}{l}{\textit{ with PMS}} \\

16 &0.7& 75 &43.3& 2 & 5 & 2.7 &1.4& 13.1 & 10.8 & 9.6 & 17.3 &1.31 \\
17 &0.9& 75 & 43.3 &  2 & 5 & 1.9 &1.4& 11.6 & 13.8  & 6.6 & 26.0 & 0.90\\
18 &0.9& 75 & 43.3 &3 & 5 & 5.0 &1.4& 12.7 & 17.8 & 10.1 &23.0 & 1.36\\
19 &0.9& 100 & 57.7 & 2 & 6  & 1.6 &1.4& 15.9 & 16.5 & 8.9 & 30.8 & 1.34\\ 
20 &0.9& 100 & 57.7 &  2 & 5  & 3.0 &1.4& 12.7 & 13.9 & 8.3 & 23.4 & 1.30\\
21 &0.9& 100 & 57.7 & 3 & 5  & 7.0 &1.4& 16.9 & 17.7 & 14.8 & 21.0 & 1.37 \\
22 &0.9& 100 & 57.7 &  4 & 5  & 8.5 &1.4& 20.2 & 21.2 & 19.0 & 24.1 & 1.41\\ 
23 &0.9& 130 & 75.0 & 2 & 5  & 3.4 &1.4& 21.4& 14.7 & 16.1 & 22.4 & 1.22\\
24 &0.9& 131 & 75.0 & 2 & 6 & 2.1 &1.4& 22.1 & 16.5 & 14.5 & 29.4 & 1.38 \\
25 &0.9& 140 & 80.0 & 2 & 7 & 1.6 &1.4& 26.4 &18.0 & 16.9 & 33.7 & 1.39 \\ 
26 &0.9& 173 & 100.0 & 2 & 5  & 4.1 &1.4& 33.0 & 14.9 & 27.2 & 20.6 & 1.42\\
27 &0.9& 174 & 100.0 & 2 & 6  & 3.6 &1.4& 22.4 & 17.2 &16.2 & 26.2 & 1.31\\
28 &0.9& 175 & 100.0 & 2 & 7  & 1.6 &1.4& 36.5 & 17.9 & 24.9 & 32.0 & 1.27 \\
\hline
\multicolumn{11}{l}{\textit{ with PMS and $ \alpha\approx1 $ at $t_\alpha$}} \\
29 &0.9& 175 & 100.0 & 2 & 7 & 2.3 &1.1& 25.1 & 18.6 &16.2 &32.8 & 1.14\\
\bf30 &\bf0.9& \bf175 & \bf100.0 & \bf2 & \bf7  & \bf2.5 &\bf1.0& \bf20.3 & \bf19.2 & \bf12.5 & \bf33.7 & \bf0.95 \\
\hline
\multicolumn{10}{l}{\textit{ with PMS and eccentric orbit in the MW($ e\approx0.6 $)}} \\
\bf31 &\bf0.9& \bf140 & \bf80.0 & \bf2 & \bf7 & \bf1.6 &\bf1.4& \bf26.4 & \bf18.0 & \bf11.8 & \bf 32.5  & \bf1.21\\
32 &0.9 & 175 & 100.0 & 2 & 7 & 2.3 &1.1& 25.1 & 18.6 & 12.3 & 31.6 & 1.03\\ 

\hline
OBS& && & & & && & & $12.0$& $35.6$& $ 1.27$ \\
& && & & & & && & $(\pm 1.0)$& $(\pm 0.6)$& $(\pm 0.44)$ \\
\hline
  \end{tabular}
\end{table*}
%

%
%
\begin{figure}
  
\centering
   \includegraphics[width=8CM]{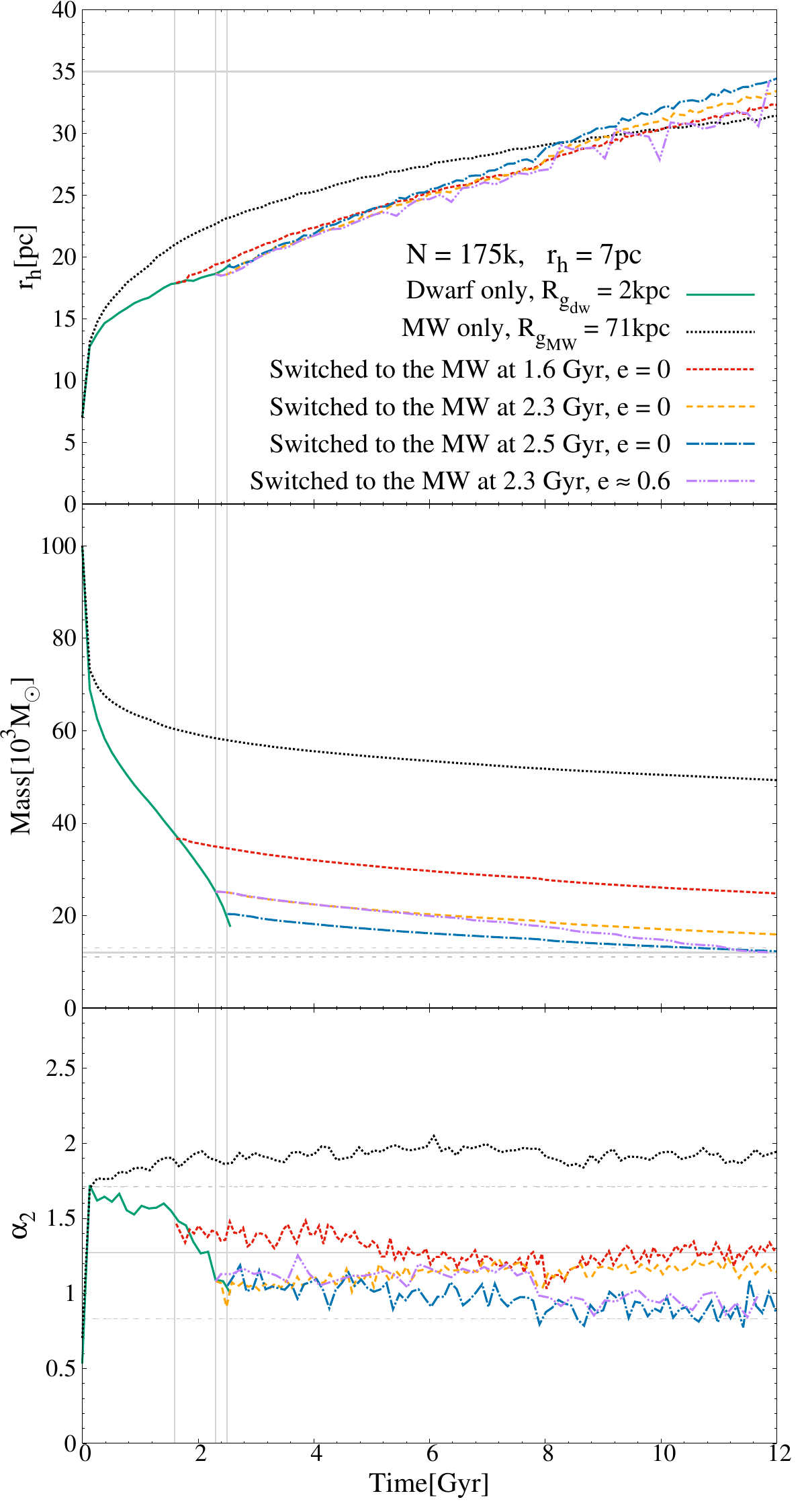}

\caption{Evolution of the half-mass radius (top), the total mass (middle) and $\alpha_2$ 
(bottom) of the cluster with $N=1.75\times10^5$ particles and an initial 3D post-gas expulsion half-mass radius of 7 pc. In this model, the degree of mass segregation was chosen as $S=0.9$. The cluster evolves around a point mass dwarf galaxy, with $ M=1.0\times10^9M_\odot$ at an orbital distance $R_{\{g,dw\}}=2 \rm kpc$ (green solid lines), entirely on a circular orbit around the Milky Way at $R_{\{g,MW\}}=71 \rm kpc$ (black dotted lines), and has been switched from the dwarf galaxy potential to the Milky Way potential in four different models: on a circular orbit at $t=1.6 \rm Gyr$ (red dashed lines); on a circular orbit at $t=2.3 \rm Gyr$ (orange dashed lines); on a circular orbit at $t=2.5 \rm Gyr$ (blue dash-dotted lines), and on an eccentric orbit with $ e\approx0.6 $ and $ R_a =71 \rm kpc$, at $t=2.3 \rm Gyr$ (purple dash-dotted lines).  Vertical lines show the corresponding $t_{\alpha}$ to each model (Table 1). The horizontal lines show the observed values.}

\label{fig:3}

\end{figure}
%
%
\begin{figure}

    \centering
   \includegraphics[width=8CM]{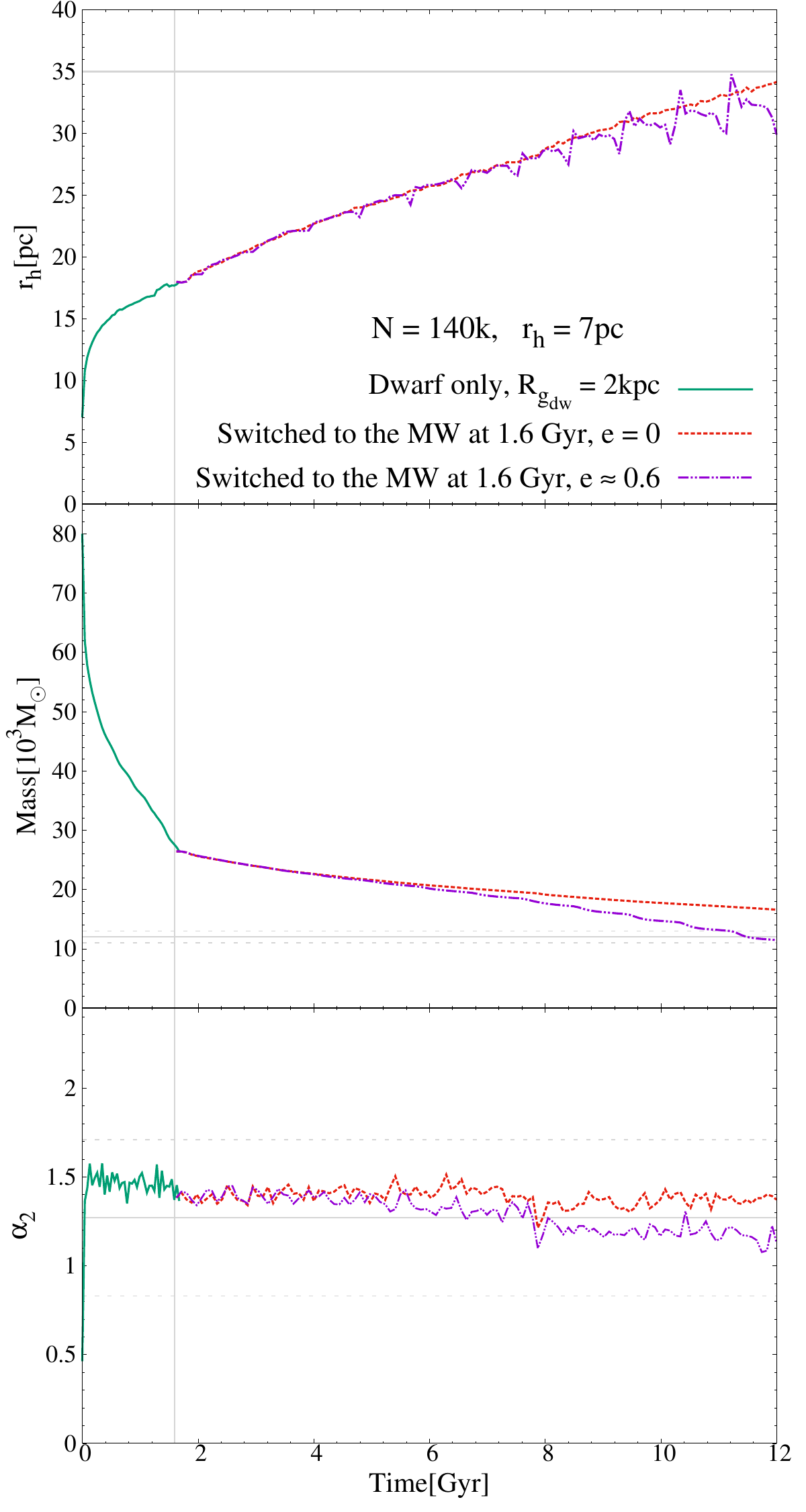}

\caption{Evolution of the half-mass radius (top), the total mass (middle) and $\alpha_2$ 
(bottom) of the cluster with $N=1.40\times10^5$ particles and the initial 3D post-gas expulsion half-mass radius of 7 pc. In this model, the degree of mass segregation was chosen as $S=0.9$. The cluster evolves around a point mass dwarf galaxy, with $ M=1.0\times10^9M_\odot$ in an orbital distance $R_{\{g,dw\}}=2 \rm kpc$ (green solid lines), and has been switched from dwarf galaxy potential to the Milky Way potential in two different models: place into the Milky Way, in a circular orbit at $t=1.6 \rm Gyr$ (red dashed lines) and in an eccentric orbit with $ e\approx0.6 $ and $ R_a =71 \rm kpc$, at $t=1.6 \rm Gyr$ (purple dash-dotted lines). Vertical line shows the corresponding $t_{\alpha}$ (Table 1). The horizontal lines show the observed values.}

\label{fig:4}
\end{figure}

\subsection{Models with PMS}    \label{sec:PMS} 

As shown in Section~\ref{sec:withoutPMS}, the initially non-segregated clusters with a canonical  IMF could not reproduce the observed half-mass radius of Palomar 14. To understand whether PMS helps to reconcile the inconsistency between observations and simulations, we calculated several models starting with PMS (models 16-32, Table~\ref{table_1}). 

Fig.~\ref{fig:2}  compares the evolution of the 3D half-mass radius, total mass, and MF-slope for three models with the same initial number of particles ($ N=7.5\times10^4 $) and initial half-mass radius $(r_h=5 \rm pc)$, but different values of PMS: $S = 0$ (Model 7), $S=0.7$ (Model 16) and $S=0.9$ (Model 17). All clusters evolve at the same orbital distance around the dwarf galaxy ($R_{\{g,dw\}}=2$ kpc). We find that the final half-mass radius of a modeled cluster is very sensitive to its degree of PMS. For example, for models initialized with  $10^5 M_\odot$, the final radius changes from about 10 to 25 pc when we change the initial PMS from $S=0$ to $0.9$ (Fig. \ref{fig:2}).  The models with PMS undergo a stronger initial expansion, owing to the early impulsive mass-loss associated with stellar evolution which is now happening preferentially deep inside the cluster core, reaching larger half-mass radii compared to models starting without PMS. In the model without PMS (model 7 shown as a blue dashed line), the cluster reaches $\alpha=1.4$ at $t_{sw}\approx6.2$ Gyr, while in the fully segregated model (model 17 shown as a brown dashed line), it occurs at $t_{sw}\approx 1.9$ Gyr, with a larger final half-mass radius of $ r_{h,sw} \approx 13.8$ pc. We conclude that when the cluster is initially mass segregated, the flattening of the MF occurs faster, and the cluster is larger than models without PMS. After entering the outer halo area of the MW, the MF-slope continues to change slowly, while the cluster expands further for models with a higher $S$-value and reaches larger radii. 


In order to find the most likely initial conditions for Pal 14 we have computed a grid of 12 models with the total number of stars between 75000 and 175000, and with different initial half-mass radii in the range, $r_h = 5 - 7$ pc (we implicitly assume the cluster is born with a birth radius \citep{Marks2012}, and gas expulsion expands this to the initial radius used here. Among the models with PMS,  only models $24-28$ in Table~\ref{table_1} come close to the observed parameters. But none of them can simultaneously reproduce the observed present-day mass and half-mass radius of Pal 14 even when accounting for observational and statistical errors.

Fig.~\ref{fig:3} shows the evolution of the MF-slope (bottom panel), the total mass (middle panel), and the half-mass radius (top panel) for model numbers 28, 29, 30, and 32 in Table~\ref{table_1}. A similar model but evolving on a circular orbit with a Galactocentric radius of 71 kpc is considered for comparison (black dotted line). As can be seen in the bottom panel of Fig.~\ref{fig:3}, if a cluster only evolves within the Milky Way, the MF-slope changes slowly. Similarly, in the models that switched to the Galaxy potential on a circular orbit, the MF-slope changes slowly (red and orange dashed lines, and blue dash-dotted line).  

According to Table~\ref{table_1}, in Model 30 we assume the cluster has evolved around the dwarf galaxy for $2.5$ Gyr (until its MF-slope reaches $\alpha_2=1$). After this time, the cluster has evolved for another $9$ Gyr in the Milky Way potential. The cluster parameters at $t=11.5$ Gyr are: $ M_f\approx12500 M_\odot $, $ r_{h,f} = 33.7$pc and $ \alpha_{2f} \approx0.95 $ (Fig~\ref{fig:3}, blue dash-dotted lines). Therefore, this model can reproduce all the present-day parameters of Pal 14 better than other models. 

\subsubsection{Evolution on eccentric orbit}

In this section, we aim to perform some N-body models orbiting on eccentric orbits to determine whether it is possible to find initial conditions with PMS in this new scenario involving MW-captured GCs.  Since star clusters lose more mass during pericentric passages on eccentric orbits, and undergo stronger expansion due to the weaker tidal fields at larger Galactic radii \citep{Baumgardt2003, Webb2014}, an eccentric cluster orbit might have influenced the evolution of Pal 14, as it could have had a much smaller initial size and significantly higher mass initially.



To investigate the effects of eccentricity, we repeat models 25 and 29  in Table~ \ref{table_1}  by changing their orbits from circular to eccentric with $ e\approx0.6 $, $ R_a\approx 71$  kpc and $ R_p\approx 20$ kpc. The purple dash-dotted lines in Fig~\ref{fig:3} and Fig~\ref{fig:4} show the evolution of cluster parameters of the clusters moving on eccentric orbits. In both models, the effect of the eccentric orbit on the MF-slope and half-mass radius is insignificant, while their final masses differ. Comparing models 25 and 31 in Table~\ref{table_1}, one can see that the cluster that moves on an elliptical orbit better matches the observations. This also applies to models 29 and 32.

\section{Summary and Conclusions}

Pal 14 is a Galactic diffuse GC in the outer halo of the MW with a two-body relaxation time of the order of a Hubble time and contains a stellar mass function depleted of low-mass stars. This paper is a follow-up to \cite{Zonoozi2011} in which we modeled the dynamical evolution of this GC over its entire lifetime on a star-by-star basis using the state-of-the-art direct N-body code \textsc{nbody6}. As a possible solution for the observed flattened stellar MF of a remote halo, extended GCs such as Pal 14, we have suggested that this object was formed as a star cluster in a dwarf galaxy satellite later acquired by the MW.

In this paper, we have performed a large grid of simulations of multi-mass star clusters that start in a dwarf galaxy potential and are then instantaneously switched to a Milky Way potential to model Pal 14 over its entire lifetime on a star-by-star basis to find likely initial conditions that reproduce its observed mass, half-light radius, and slope of the stellar mass function of Pal 14. The important question we aim to answer is whether we still need PMS to explain the flattening of the MF in the Pal 14 in the context of it possibly being an acquired GC. 

We have chosen the total mass, the half-mass radius of the cluster, and the orbital Galactocentric radius of the cluster in the dwarf galaxy as three free parameters, and by varying the initial conditions, we obtained the most likely initial conditions that reproduce its observed mass, the half-mass radius, and the MF-slope which was found to be significantly shallower than in most GCs. In all models, we assume that Pal 14 was born in a dwarf galaxy and evolved on a circular orbit around a dwarf galaxy with a strong tidal field until the MF-slope approaches the measured value, then switched to the weaker tidal field of the Milky Way potential on a circular orbit. The main difficulty in our calculations is finding initial models that simultaneously reproduce all these structural parameters of Pal 14.   

We found that the present-day parameters of modeled clusters qualitatively depend on the time that the cluster is taken from the dwarf galaxy. Our work suggests that the initial conditions of Pal 14 must have been a post-gas-expulsion half-mass radius of about $r_h=7$ pc and a mass of about $M=$ 80-100k M$_{\odot}$.  By increasing the duration of the GC's presence in the dwarf galaxy, it is possible to start with a higher initial mass.  

Since the GAIA DR3 data of Pal 14 orbital parameters suggest that its orbit is highly eccentric, to determine if such an eccentric orbit would affect the cluster's post-capture evolution, we conducted tests and found that models without PMS are not able to simultaneously reproduce all observed data. In a general study, we calculated additional initial models with PMS that evolved on an eccentric orbit within the Milky Way after switching from the circular orbit in the dwarf galaxy. We discovered that clusters moving on an elliptical orbit have a better match with the observations. We concluded the need for a high degree of PMS among the cluster stars is crucial for all scenarios we have considered.

\section*{Data availability}
The data underlying this article are available in the article.


\bibliography{references}{}
\bibliographystyle{aasjournal}

\end{document}